\documentclass[useAMS,usenatbib]{mn2e}

\usepackage{graphicx}
\usepackage{amssymb}

\title[A 850 $\mu$m Survey of  Discs in IC~348]{A SCUBA-2 850 $\mu$m Survey of Protoplanetary Discs in the  IC~348 Cluster}
\author[L. Cieza et al.]{\Large L. Cieza$^{1,2}$, J. Williams$^3$,  E. Kourkchi$^3$, S. Andrews$^4$,  S.  Casassus$^{2,5}$,  S. Graves$^6$, and  M. Schreiber$^{2,7}$
\\
$^{1}$N\'ucleo de Astronom\'ia,  Facultad de Ingenier{\'i}a, Universidad Diego Portales,  Av. Ejercito 441, Santiago, Chile\\
$^{2}$Millenium Nucleus ``Protoplanetary Disks in ALMA Early Science", Chile \\
$^{3}$Institute for Astronomy, University of Hawaii at Manoa, Honolulu, HI 96822, USA\\
$^{4}$Harvard-Smithsonian Center for Astrophysics, 60 Garden Street, Cambridge, MA 02138, USA\\
$^{5}$Departamento de Astronom\'ia, Universidad de Chile, Casilla 36-D Santiago, Chile\\
$^6$Joint Astronomy Centre, 660 N. Aohoku Place, Hilo, HI 96720, USA \\
$^{7}$Instituto de Astronom\'ia y Astrof\'isica, Universidad de Valpara\'iso, Avda. Gran Breta–a 1111, Valpara'so, Chile
}
\begin{document}


\pagerange{\pageref{firstpage}--\pageref{lastpage}} \pubyear{2002}

\maketitle

\label{firstpage}

\begin{abstract}
We present  850 $\mu$m observations of  the 2--3 Myr cluster IC~348 in the Perseus molecular cloud using the SCUBA-2 camera on the James Clerk Maxwell Telescope. 
Our SCUBA-2 map has a diameter of  30$\arcmin$ and contains $\sim$370 cluster members, including $\sim$200 objects with
IR excesses.  We detect a total of 13 discs.  Assuming standard dust properties and a gas to dust mass ratio of 100,  we derive disc masses  ranging from 1.5 to 16 M$_{JUP}$.   We also detect  8 Class 0/I protostars.  
We find that the most massive discs (M$_D$ $>$ 3 M$_{JUP}$; 850 $\mu$m flux $>$ 10 mJy) in IC~348  tend to be transition objects according to the characteristic ``dip" in their infrared  Spectral Energy Distributions (SEDs). This trend is also seen in other regions. We speculate that this could be an initial conditions effect (e.g., more massive discs tend to form giant planets that result in transition disc SEDs) and/or a disc evolution effect (the formation of one or more  massive planets  results in both a transition disc SED and a reduction of the accretion rate, increasing the lifetime of the outer disc). 
A stacking analysis of the discs that remain undetected in our SCUBA-2 observations suggests that their median  850 $\mu$m flux should be 
$\lesssim$ 1 mJy, corresponding to a disc mass  $\lesssim$0.3 M$_{JUP}$ (gas plus dust) or $\lesssim$1 M$_{\oplus}$ of dust.  
While the available  data are not deep enough to allow a meaningful comparison of the disc luminosity functions between IC~348 and other young stellar clusters,  our results imply
that disc masses exceeding the Minimum Mass Solar Nebula are very rare ($\lesssim$1$\%$) at the age of IC~348, especially around very low-mass stars. 
\end{abstract}


\section{Introduction}

Over the last dacade, \emph{Spitzer} has tremendously enriched our knowledge of the structure and
evolution of circumstellar discs, the birth sites of planets.  From  \emph{Spitzer} studies of molecular clouds and young stellar clusters
much has been learned about the infrared (IR) properties of pre-main-sequence (PMS) stars as a function of stellar age and mass (Carpenter et al. 2006; Allen et al. 2007;   Hernandez 2007; Kim et al. 2013). 
\emph{Spitzer} observations have also revealed the diversity of radial  structures that discs can present (e.g., holes and gaps) and the many 
evolutionary paths that discs can follow (Currie et al. 2011; Williams $\&$ Cieza, 2011; Espaillat et al. 2014). 

While most protoplanetary discs remain optically thick in the IR, they become optically thin at longer wavelengths. 
Continuum (sub)millimeter observations are hence sensitive to dust properties (total dust masses and grain size distributions) and are highly complementary to IR data.
The (sub)millimeter luminosities of protoplanetary discs with almost identical IR Spectral Energy Distributions (SEDs) can differ by up to two orders of magnitude (Cieza et al. 2008; 2010) 
implying drastically different disc masses and/or grain-size distributions; therefore, understanding disc evolution requires both IR and submillimeter observations. 

Unfortunately, submillimeter surveys  of discs in molecular clouds and young stellar clusters clearly lag behind their IR counterparts. 
While complete IR censuses of discs exist for tens of regions,  there is only one region in which the entire disc population has been observed at (sub)millimeter wavelengths with
enough depth to detect the majority of the IR-detected discs:  the 1-2 Myr Taurus molecular cloud (Andrews et al. 2013). In Taurus, the  typical mass of a disc around a 1 M$_{\odot}$ star is $\sim$4 M$_{JUP}$ and 
disc masses are roughly proportional to stellar mass (although there is a large dispersion for any given stellar  mass).
Our recent 850 $\mu$m SCUBA-2 observations of  $\sigma$-Orionis (age $\sim$3-5 Myr; members $\sim$300)  show that massive  discs ($>$ 5 M$_{JUP}$) are rare ($\sim$3$\%$ ) around members of this older  cluster  (Williams et al. 2013).  Even though  those observations were insensitive to less massive discs, a stacking analysis of the undetected sources suggests the mean mass of the IR-identified primordial discs (N $\sim$80 ) is of the order of 0.5 M$_{JUP}$,  indicating a rapid decline in the amount of raw material that is available for planet formation
between the age of Taurus and the age of  $\sigma$-Orionis.

Here we present 850~$\mu$m SCUBA-2  observations of the IC~348 cluster,  which is situated at the eastern edge of the Perseus molecular cloud 
complex,  at a distance of 320 pc (Herbig 1998). 
IC~348 is a rich cluster partially embedded in natal gas and dust. This cluster has been extensively studied to determine its membership and
contains $>$350 spectroscopically confirmed members (Luhman et al. 1998, 2003, 2005; Muench et al.  2007) with masses ranging from 0.02 to 5 M$_{\odot}$ and an estimated mean age of 2-3 Myr (Herbig 1998). 
This age range places IC~348 right in between Taurus and $\sigma$-Ori, which is consistent
with its intermediate IR disc fraction of 50$\%$ (Lada et al. 2006), to be compared with 63$\%$ for Taurus (Hartmann et al. 2005) and 27$\%$ for $\sigma$-Ori (Hernandez et al. 2007).
IC~348 is slightly closer than $\sigma$-Orionis (320 vs 420 pc) and its 850 $\mu$m data is slightly deeper (2.7 mJy vs 2.9 mJy rms median noise). 
The resulting disc mass sensitivity is therefore a factor of  $\sim$2.0 better in IC~348 than in  $\sigma$-Ori ($\sim$2.5 vs 5.0 M$_{JUP}$)

IC~348 has already been observed at 1.3 mm with the Submillimeter Array (SMA, Lee et al. 2011). 
The 22 SMA fields from that study are each 55$\arcsec$ in diameter and in total include 85 cluster members, about 25$\%$ of the entire population. 
Nine young stellar objects were detected by the SMA with disc masses larger than $\sim$2 M$_{JUP}$. 
The 850 $\mu$m SCUBA-2 map presented here has a similar disc mass sensitivity, but covers all $\sim$370 cluster members. 
The SCUBA-2 observations are described in Section 2. The main results are shown in Section 3. 
In Section 4, we discuss  the correlations between the (sub)millimeter  and  the IR properties of the targets and place our results in the general context of disc evolution
and planet formation. Our main conclusions and a summary of the paper are presented in Section 5.

\section[]{Observations}

The IC~348 cluster was observed with SCUBA-2 using the ``pong-900" mapping mode 
resulting in a circular region with diameter of about 30$\arcmin$. 
The center of the observed field is RA=03h:44m:34s, DEC=+32d:07m:48s and the observed area mostly covers the prominent star forming filaments of IC~348 at the north east of the Perseus complex.
The data were taken between October 2011 and October 2013 in queue mode over several 
observing runs (program IDs:   M11BH29B, M12AH29B, M12BH10C,  and  M13BH12A).
The observations were performed under  JCMT band-3  weather, 
which is defined based on the zenith optical depth at  225GHz ($\tau_{225}$) lying between 0.08 and 0.12.
The actual  $\tau_{225}$  value for each night is listed in Table 1. 
The total on-source integration time amounts to 22.4 hs. 

The pong-900 mapping mode provides fairly uniform sensitivity within a circular
region of 20$\arcmin$ in diameter. The useable area extends to a diameter of more
than 25$\arcmin$ in this mode,  but with larger noise toward the edges. 
For data reduction, the Dynamic Iterative Map Maker in the STARLINK/SMURF software package (Chapin et al. 2013) was used at the Joint Astronomy Centre. Due to the existence of
extended emission at 850 $\mu$m  and because we were most interested in discs around stars, which
would be point sources at the 15$\arcsec$ SCUBA-2 resolution, we used the blank field configuration for
the map maker which is usually used for deep field
extragalactic surveys.

The data were reduced night by night and then co-added with the weights proportional to the 
inverse square of the map noise. The flux density scale was estimated using the usual 
observations of the bright point sources, most commonly Uranus, Mars, and protoplanetary nebulae CRL 618 and CRL 2688. 
We reduced these sources with the same (blank field) map maker configuration file to derive the 
flux calibration factor (FCF). Following this procedure, we found FCF = 715 $\pm$ 70 Jy pW$^{-1}$ beam$^{-1}$and
therefore we multiplied the co-added map by this mean value.

\begin{figure*}
\includegraphics[width=210mm]{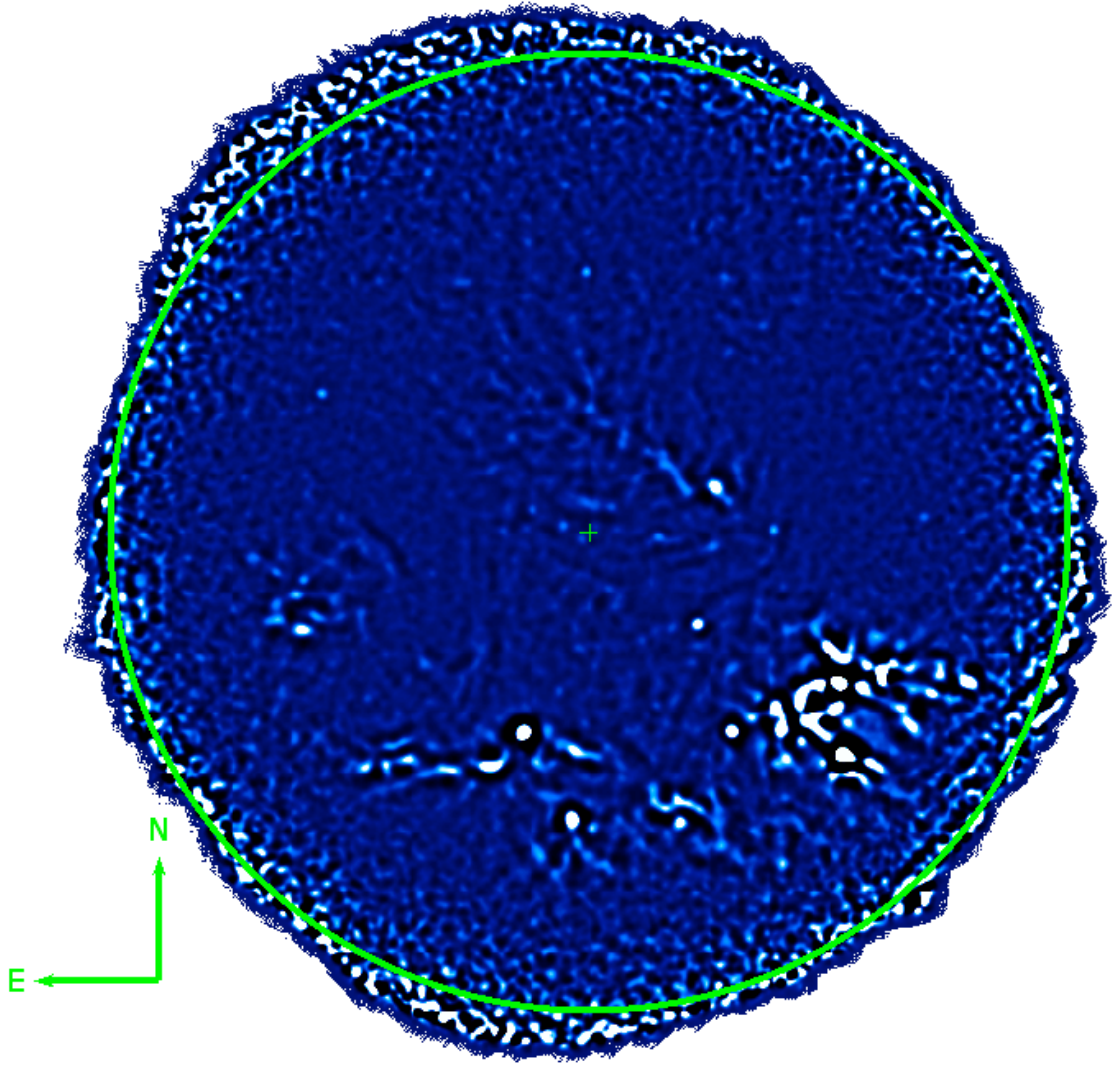}
\caption{Our 850 $\mu$m SCUBA-2 map of the IC~348 cluster.  The center of the map is indicated by  a ``+"  and corresponds to RA=03:04m34s, Dec=+32d07m48s. 
As a scale reference, the green circle has a diameter of 30$\arcmin$. The noise is more or less uniform in the inner 25$\arcmin$ region and degrades toward the edges. 
Significant extended emission from the molecular cloud can be seen in the south and southwest of the map. 
}
\label{fig:map}
\end{figure*}

At the end of the data processing, a  matched beam filter was applied to smooth the
data and maximize the sensitivity to point sources. Figure~\ref{fig:map} shows the final 850 $\mu$m map with 3$\arcsec$ pixels we use for our study.
Significant extended emission from the molecular cloud is seen toward the south and south-west of the map.
SCUBA-2 observes at 850 and 450 $\mu$m simultaneously, but we do not use 450 $\mu$m data in our analysis because our program does not meet the stringent weather requirements for
sensitive 450 $\mu$m  observations ($\tau_{225}$ $<$ 0.05). 

\section[]{Results}

\subsection{Detections}

\begin{figure*}
\includegraphics[width=205mm]{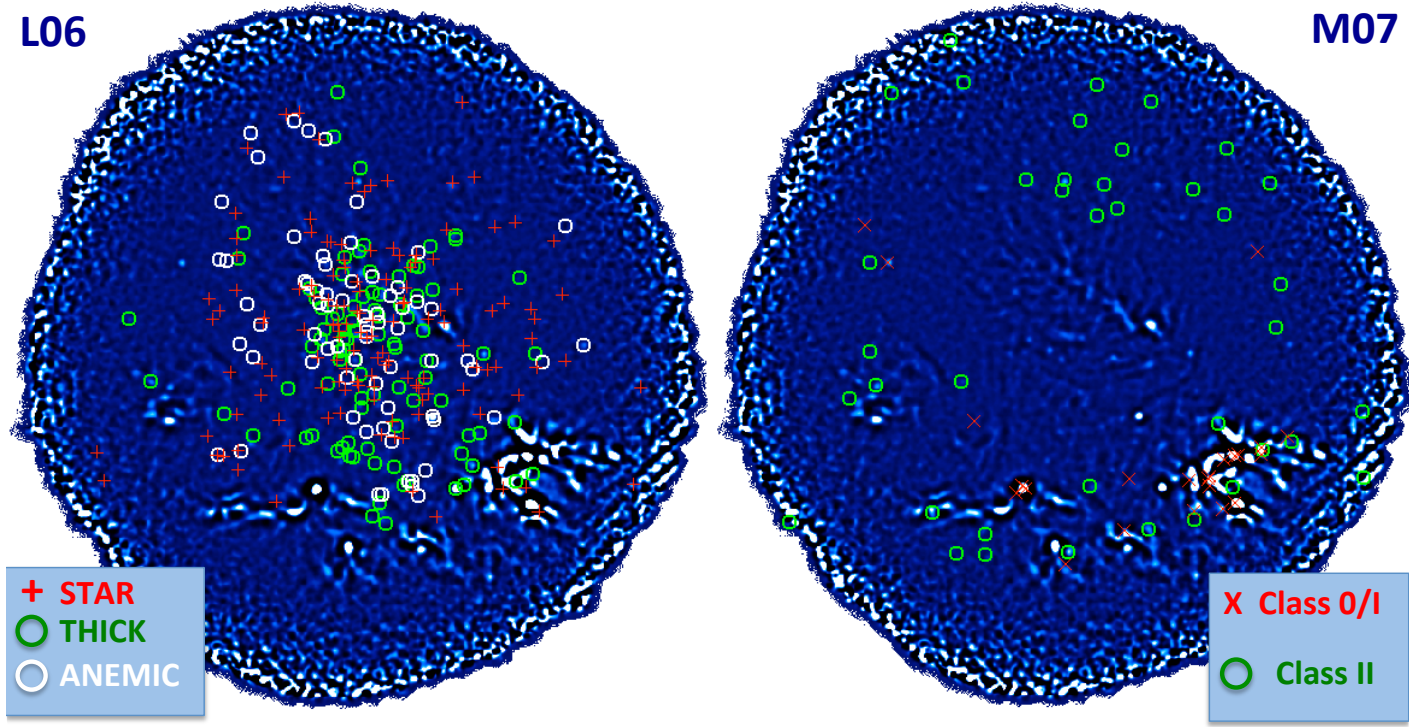}
\caption{The spatial distributions of the different types of sources listed by Lada et al. (2006, left panel) and Muench et al. (2007, right panel). The sources from Lada et al. are distributed toward the center of the map, while
the sources from Muench at al.  tend to be located closer to the edges.}
\label{fig:L06M07}
\end{figure*}

We base our 850 $\mu$m analysis on the \emph{Spitzer} sources studied by  Lada et al. (2006, L06 hereafter) and Muench et al.  (2007, M07 hereafter).  
Our SCUBA-2 map is $\sim$30$\arcmin$ in diameter and contains the 307 cluster members studied by L06 and 62 of the
66 additional members studied by M07.  As seen in Figure~\ref{fig:L06M07}, the L06 sources concentrate toward the center of the map, while the M07 sources tend to lie toward the edges of the map and along the extended cloud emission seen in the south and south-west of the map.
The L06 sources are confirmed cluster  members from the spectroscopic study by Luhman et al. (2003), while the M07 sources are additional \emph{Spitzer}-selected objects with membership confirmation from followup  ground-based spectroscopy. As a result, the latter group tend to be more embedded sources away from the center of the SCUBA-2 map. This translates into higher noises in our 850 $\mu$m map. The median and mean noises of the L06 targets are 2.6 mJy and 4.2 mJy, respectively, while the median and mean noises of the M07 sources are 5.5 and 24.7 mJy, respectively.  Over the combined sample, L06 plus M07, the median and mean noises are 2.7 mJy and 7.6 mJy, respectively. 
The large difference between the median and mean noises in the samples implies that the noise is far from homogenous across the map and strongly dependent on the underlying cloud emission. 

Due to the variable background noise of our SCUBA-2 map, we adopt a very localized approach  to perform our photometry and derive the fluxes from a single pixel (3$\arcsec$ on a side and in units of mJy per beam), corresponding to the nominal location of the sources (based on their \emph{Spitzer} coordinates). In the context of PSF-fitting photometry, this is analogous to scaling a point-spread-function (PSF) to the central pixel, instead of allowing for a small error in position and fitting the entire PSF, which would increase the chances of confusing a local peak in the background with a real source. 
To estimate the background noise,  we use a rather small annulus with an inner and outer radius of 15$\arcsec$ and 30$\arcsec$, respectively (to obtain a localized measurement of the noise).

\begin{figure*}
\includegraphics[width=205mm]{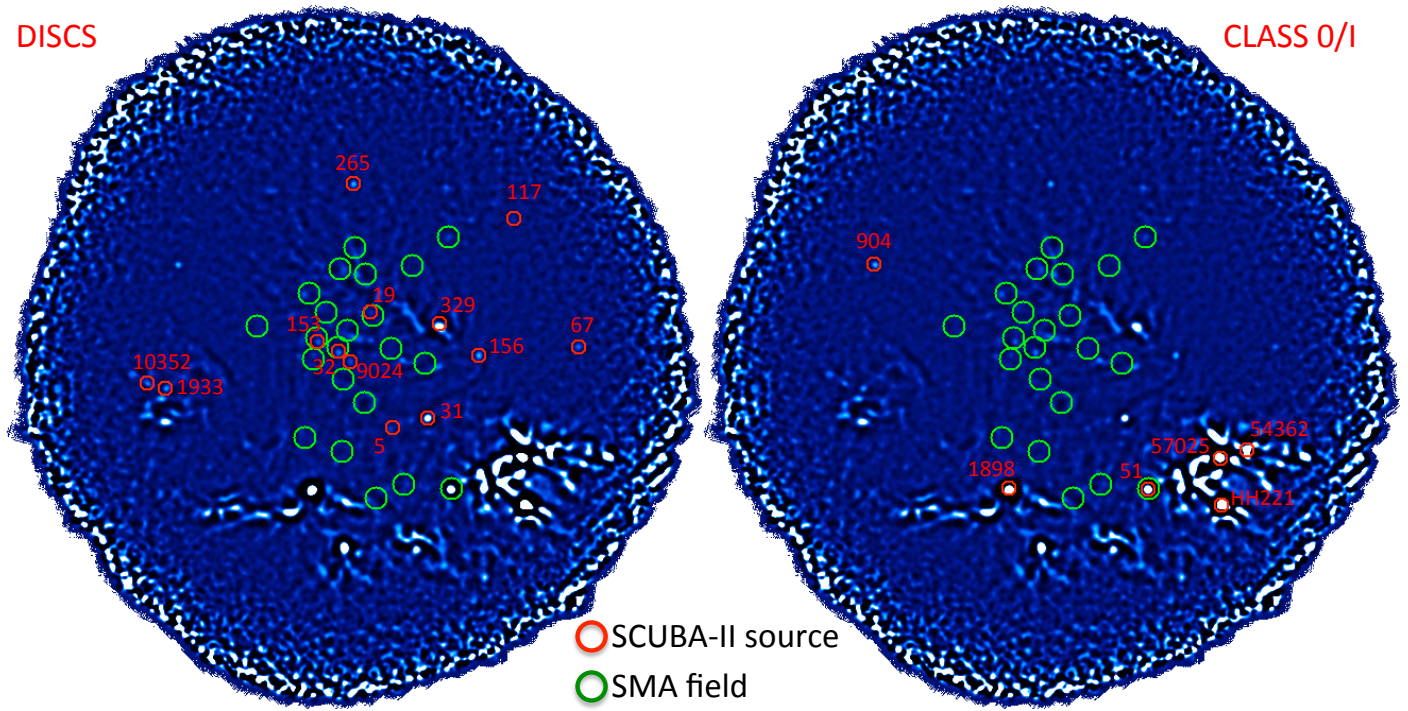}
\caption{The locations of our SCUBA-2 detections. The discs (anemic, thick and Class-II sources) are shown in the left panel. Class 0/I protostars are shown in the right panel. 
Four of our SCUBA-2 sources, \#19, 32, 51, and 153,  fall in the SMA fields observed by Lee et al. (2011), shown as green circles, 55$\arcsec$ in diameter.}
\label{fig:detections}
\end{figure*}

\begin{figure*}
\includegraphics[width=205mm]{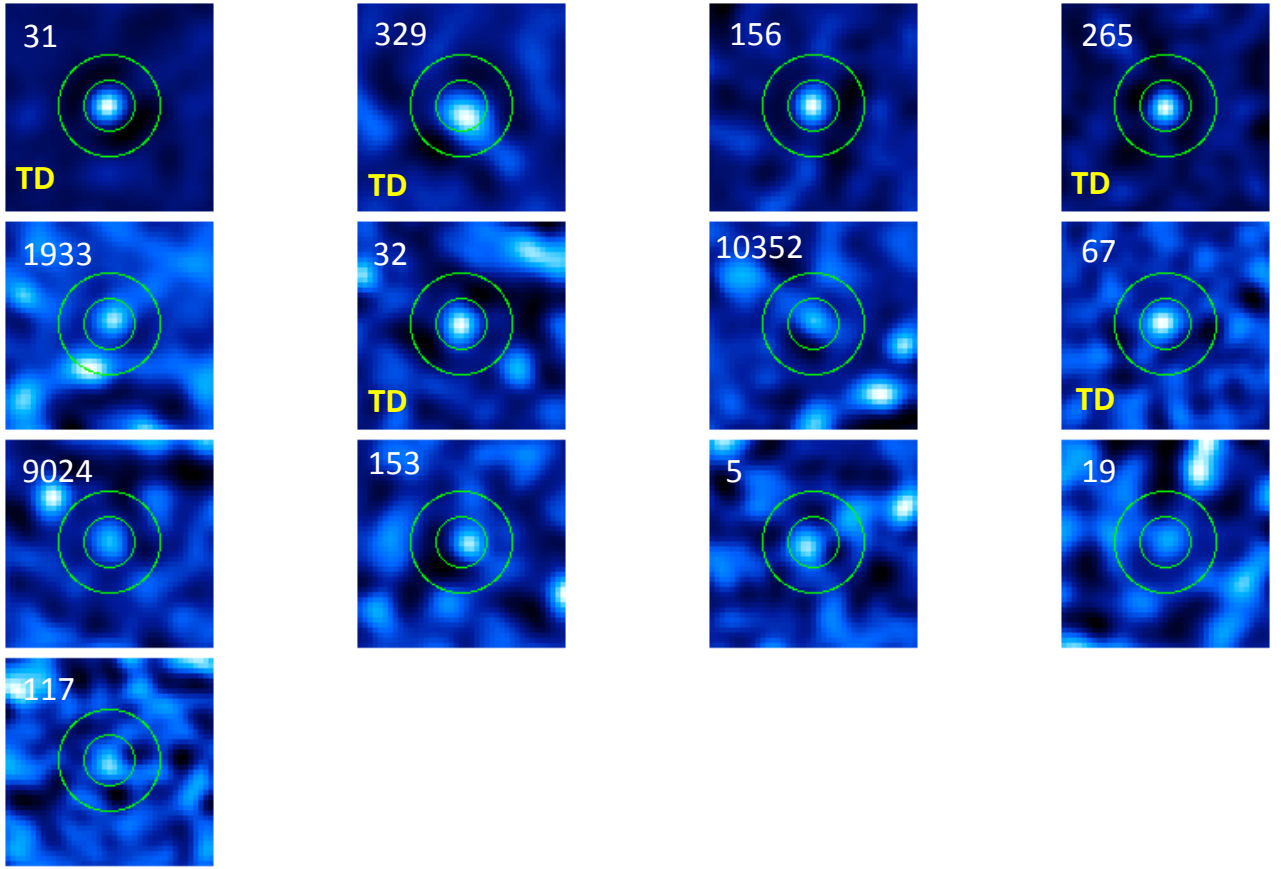}
\caption{The 13 discs detected in our SCUBA-2 map ordered by decreasing  850 $\mu$m flux.  Five objects with transition disc SEDs are labeled  ``TD" and are among the brightest eight sources. 
The green circles are 15$\arcsec$ and 30$\arcsec$ in radius. The annuli within the circles are used to estimate the local noise of the map around each source.}
\label{fig:discs}
\end{figure*}

Using this approach, we detect  21 members with signal to noise ratios $>$3.0.  Thirteen of these members were classified by L06 and M07 as 
circumstellar discs (anemic,  thick, and Class-II in their nomenclature) based on their near-IR properties, while the other eight objects are protostars (Class 0/I sources), where the 850 $\mu$m flux is likely to be dominated by their envelopes as opposed to their discs. The properties of the 13 discs are listed in Table 2
 and their  locations are shown in Figure~\ref{fig:detections} (left panel).  For identification purposes, we adopt the ID numbers used by Luhman et al. 2003, L06, and M07. Close up images of the 13 detected discs are shown in Figure~\ref{fig:discs}.

Four out of the 13 discs detected at 850 $\mu$m, sources $\#$19, 32, 51, and 153,  fall within the SMA fields observed at  1.3 mm by Lee at al. (2011), see Figure~\ref{fig:detections}.  All of them were detected by the SMA, except for $\#$19, the faintest  of the four.  This object is 5.9$\pm$1.7 mJy and thus only  marginally detected (3.5--$\sigma$) at 850 $\mu$m; however,  its SCUBA-2 detections is still consistent with the SMA non-detection, given the noise of the 1.3 mm observations (3--$\sigma$ = 2.2 mJy) and the observed 850$\mu$m to 1.3 mm flux ratios of other objects in the sample (e.g. 2.6 for source $\#$32). 
Lee et al. also detected 7 objects that were not detected in our SCUBA-2 map. Their 1.3 mm fluxes are in the $\sim$2-3 mJy range and their SCUBA-2 non-detections are also consistent with the 850 $\mu$m noise levels at their location ($\sim$2-3 mJy). These 7 objects are also listed in 
Table 2. 

\begin{figure*}
\includegraphics[width=205mm]{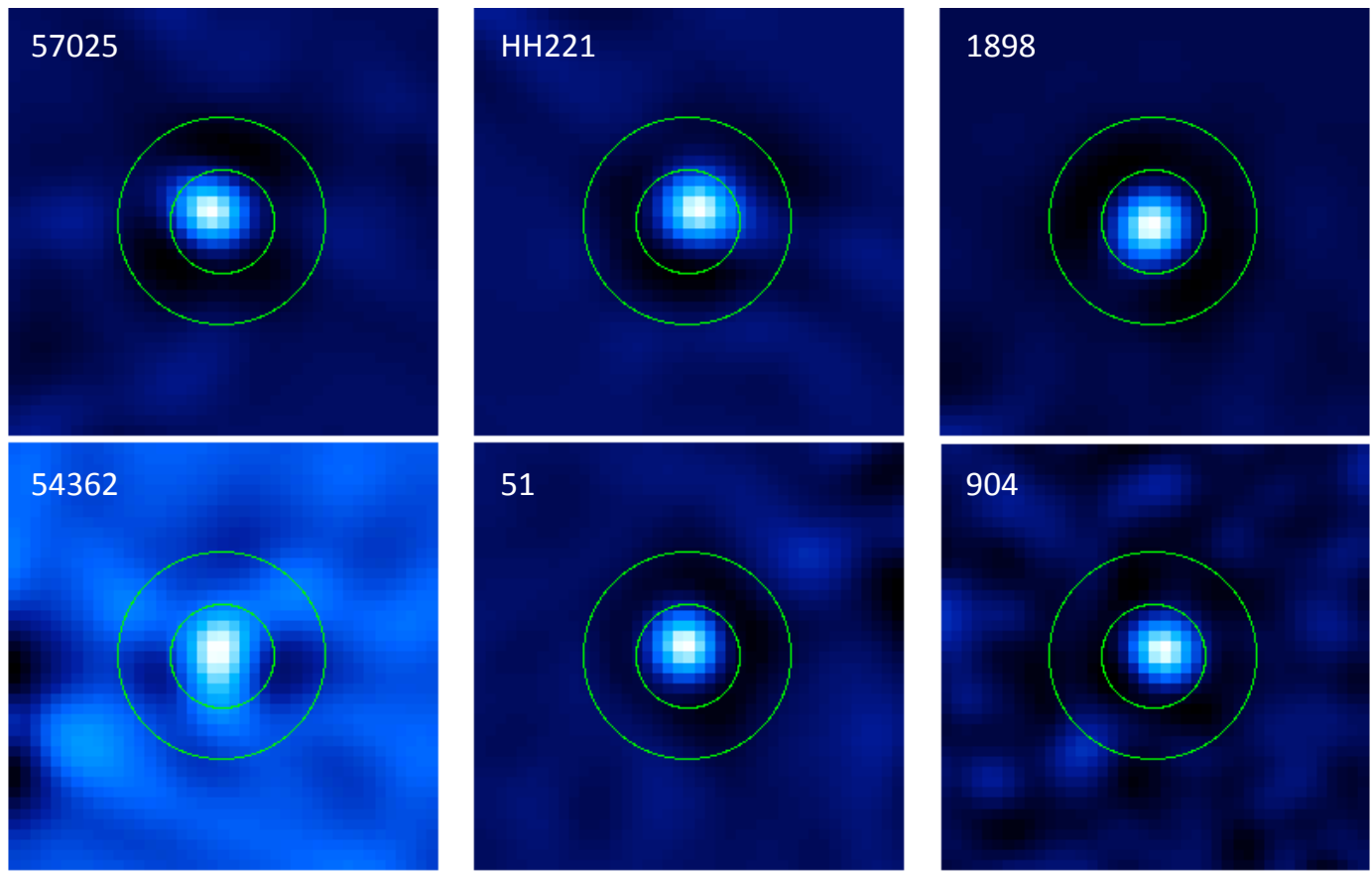}
\caption{The 8 protostars detected in our SCUBA-2  map ordered by decreasing  850 $\mu$m  flux. 
The green circles are 15$\arcsec$ and 30$\arcsec$ in radius. The annuli within the circles are used to estimate the local noise of the map around each source.}
\label{fig:proto}
\end{figure*}

In this study we focus on the disc sources in IC~348.  For completeness we also list the fluxes of the 8 detected protostars (see Table~\ref{table:proto}). Their fluxes range from $\sim$450 to 750 mJy for Class 0 sources and $\sim$20 to 200 mJy for  Class 0/I objects. Their locations are shown in Figure~\ref{fig:detections} (right panel) and their close up images in Figure~\ref{fig:proto}, but they are not further discussed in this paper. 

\subsection{Stacking of non-detections:}\label{stacking}

Even though its IR disc fraction is close to 50$\%$, the vast majority of the cluster members in IC~348 remain undetected at 850 $\mu$m. Still, it is possible to place some constraints on the typical  submillimeter
properties of the undetected discs from the statistics of the measured SCUBA-2 fluxes at the locations of the targets (i.e., by stacking the non-detections). 
L06 classified the discs in IC~348 based on the slope of the SED at IRAC\footnote{\emph{Spitzer}'s IR Array Camera} wavelengths (3.6-8.0 $\mu$m), $\alpha_{IRAC}$ = $d log \lambda F_{\lambda} / d log \lambda$,  
from a power-law, least-squares fit to the four IRAC  bands for each star.  In particular,  objects with $\alpha_{IRAC}  > -$1.8 were classified as  optically  thick discs, while objects with
$-2.56  <  \alpha_{IRAC} <  -$1.8 were dubbed ``anemic" discs.  Objects with $\alpha_{IRAC}  <  -$ 2.56 were regarded as discless.  
This classification is mostly empirical and is based on the  fact that the $\alpha_{IRAC}$ values of   PMS stars in Taurus are well separated in two groups, those with $\alpha_{IRAC}  <  -$ 2.56, in agreement with the 
predicted slope for an M0 star, and those with $\alpha_{IRAC}  > -$1.8, in agreement with the predictions for optically thick discs  extending inward to the dust sublimation radius. 
Objects with intermediate $\alpha_{IRAC} $ values were later seen in IC~348. They were  interpreted as heavily depleted or optically thin discs and grouped in the intermediate ``anemic" disc category. 
Similarly,  M07 classified the sources with $ -0.5 >  \alpha_{IRAC} >  -$1.8 as Class II and sources with   $\alpha_{IRAC}  >  -$0.5 as Class 0/I protostars. M07 does not report objects with weak excesses (anemic discs). 
For our stacking analysis, we adopt the disc classification from L06  and M07 as a simple 
metric to characterize the strength of the IR disc emission, but note some caveats. First, the  $ \alpha_{IRAC}$  classification is based on short IR wavelengths ($<$ 10 $\mu$m) and could misrepresent objects with
large inner holes that are characterized by weak IR excesses at  $\lambda < 10 \mu m$, but strong excesses at longer IR wavelengths (i.e., transition discs).   
Also, the stellar photospheres of the lowest-mass members of the cluster are faint at 8.0 $\mu$m and their photometry is quite noisy. As a result, the significance of the IR excesses of many of the
anemic discs listed by L06 is low ($<$ 3-$\sigma$) and some of the objects might be in fact discless stars (Cieza et al. 2007;  M07).
However, these caveats should only have a minor effect on the overall statistics. 

\begin{figure*}
\includegraphics[width=180mm]{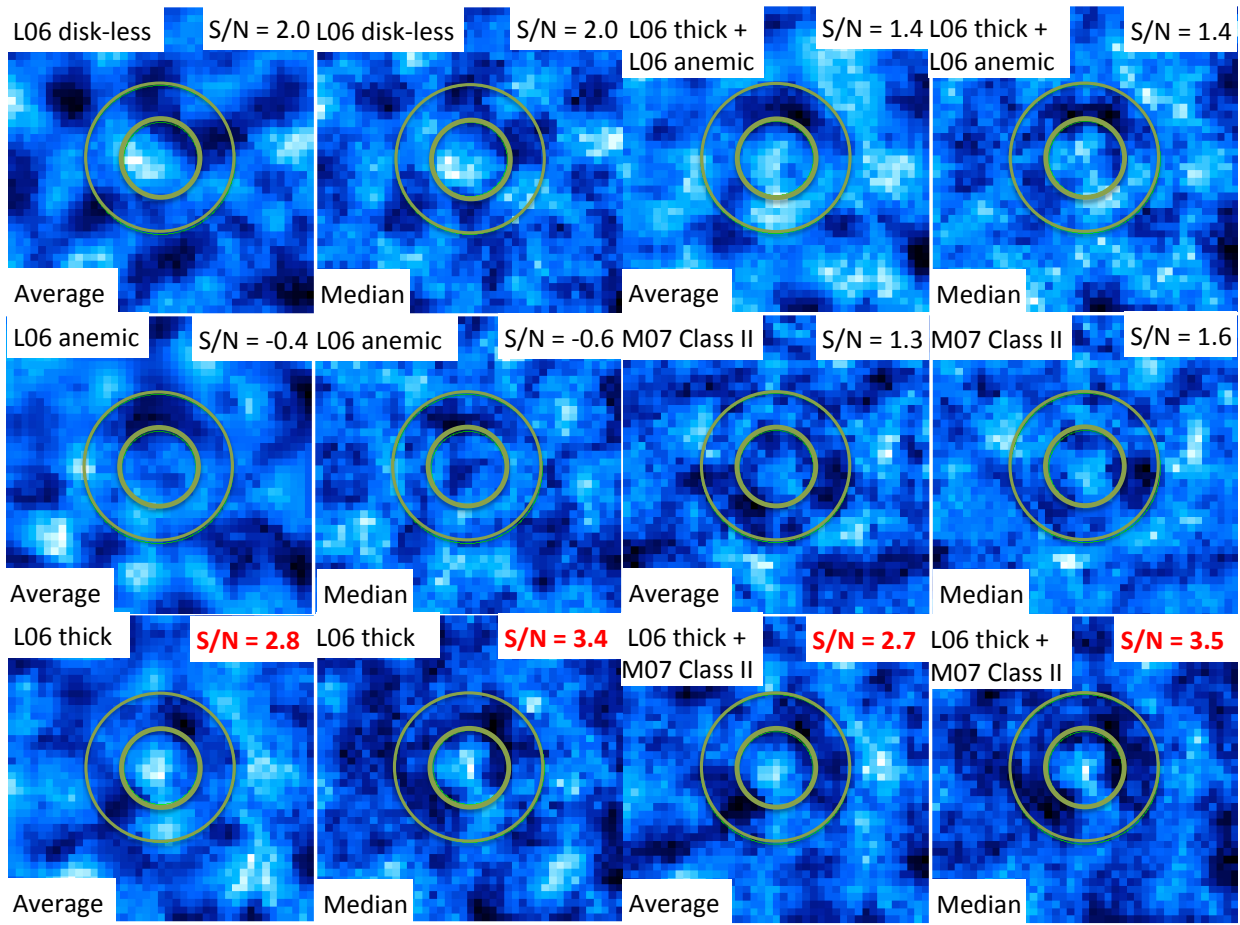}
\caption{Stacked images (average and median)
for all subsamples (discless stars, anemic discs, and thick discs from L06 and Class II  discs from M07) and some combinations of subsamples (thick discs plus anemic discs from L06 and thick discs from L06 plus Class II objects from M07).  The images are 1$\arcmin$ on the side.  Only the stacking of the  thick discs and and the thick  discs plus Class II sources show tentative detections at the 2.7 to 3.5--$\sigma$ level (bottom rows). 
 }
\label{fig:stacking}
\end{figure*}

For the stacking analysis, we produce average and median images of all subsamples (discless stars, anemic discs, and thick discs from L06 and Class II discs from M07) and some combinations of subsamples (thick discs plus anemic discs from L06 and thick discs from L06 plus Class II objects from M07). 
Before stacking,  we remove all sources that are individually detected and also mask all pixels with values larger than 10 mJy or smaller than --10  mJy in the individual images to eliminate particularly noisy regions of the mosaic. The resulting images are shown in Figure~\ref{fig:stacking}. 
We then perform the photometry as described in the previous section. As shown in Table~\ref{table:stacking} and Figure~\ref{fig:stacking}, only the stacking of the  thick  discs and the thick  discs plus the Class II sources result in possible detections at the 2.7 to 3.5--$\sigma$ level and a flux of $\sim$0.7 to 1.0 mJy. 
We regard these detections as tentative and conclude that the discs that are optically thick in the IR but remain individually undetected in our SCUBA-2 map have a median 850 $\mu$m flux $\lesssim$ 1.0 mJy.

\section[]{Discussion}

\subsection{disc masses}\label{disc-mass}

Since discs becomes optically thin at submillimeter wavelengths,  all dust grains contribute to the  emission and the total flux correlates well with 
the total dust mass.  (Sub)millimeter fluxes are thus routinely used to estimate the masses of protoplanetary discs 
using the following formula:

\begin{equation}
M_{dust}  =  \frac{F_\nu d^2}{\kappa_\nu B_\nu (T_{dust})} 
\end{equation}

where $d$ is the distance to the target,  $T$ is the dust temperature  
and  $\kappa_{\nu}$ is the dust opacity.
Adopting a distance of 320 pc, and  by making  standard  (although uncertain) assumptions about the disc temperature (T$_{dust}$ = 20 K)  and dust opacity ($\kappa_{\nu}$ = 10($\nu/$1200 GHz)cm$^2$g$^{-1}$, following Beckwith et al.  (1990), Equation 1 becomes:

\begin{equation}
M_{dust}  [M_{\oplus}] =  0.97 \times F_{850}  $(mJy)$ 
\end{equation}

From Equation 2 and adopting a gas to mass dust ratio of 100 (the canonical value used for the
 interstellar medium), the total disc mass can be estimated as follows:
 
\begin{equation}
M_{disc}  [M_{JUP}]  =  0.30 \times F_{850} $(mJy)$
\end{equation}

The disc masses  so calculated for our  SCUBA-2 detections are listed in 
Table 2. 
The disc masses for the 7 objects that were only detected by the SMA at 1.3 mm were taken directly from Lee et al.
(2011), but were calculated using the same approach.  The uncertainties in the dust opacity and gas to dust mass
ratio are significant, but the adopted values are quite standard in the literature, allowing for meaningful comparisons
to other studies. 
The detected disc have modest masses, ranging from 1.5 to 16 M$_{Jup}$.
The median noise of the IC~348 cluster members is 2.7 mJy, which translate to a 3--$\sigma$ disc mass of  2.4 M$_{JUP}$;  therefore,  our 
survey can be considered to be complete down to this level in regions free of contamination from 
cloud emission.

\subsection{Non-detections}

Equations  2  and 3 can also be used to place upper limits to the typical dust and disc masses of the 
undetected discs in IC~348. 
The $\sim$1.0 mJy  marginal detections in the  stacking analysis of the  ``Thick" and Class-II discs  translate to dust mass of $\lesssim$1.0 M$_{\oplus}$,  corresponding to a disc mass of $\lesssim$ 0.3 M$_{JUP}$ assuming a gas to dust mass ratio of 100.
Williams et al. (2013) performed a similar stacking analysis for the SCUBA-2 non-detections toward
IR-detected discs in the $\sigma$-Orionis cluster and found a positive signal (3.3-$\sigma$ significance),
consistent with an average disc mass of 0.54 M$_{JUP}$.  
Since $\sigma$-Orionis is an older region than IC~348, we speculate that the lower mass of the average disc in IC~348 is
a host mass effect:  there is a higher incidence of very low mass stars (late M-type) in IC~348 and disc masses are known
to scale with the mass of the central star (Andrews et al. 2013). 

\subsection{Correlation between submm and IR properties}

In order to investigate a possible correlation between the submillimeter and the IR properties of  the discs in IC~348, we first construct their SEDs. 
For consistency,  we adopt the IR fluxes (1.25 to 70 $\mu$m) from the catalogs of the \emph{Spitzer} Legacy Program ``From Molecular Cores to Planet-forming discs"  (Evans et al. 2009).  We also collect  available optical photometry from the literature (Luhman et al. 2003;  Littlefair et al. 2005;   Cieza $\&$ Baliber, 2006;  and Cieza et al. 2007).  We correct the optical and near-IR fluxes for extinction, adopting 
A$_V$  = 5.88 $\times $[(J-K)-(J-K)$_0$],  where (J-K)$_0$ is the intrinsic color of the star of a given spectral type, which is taken from
Kenyon $\&$ Hartman (1995).  When available in the \emph{Spitzer} archive\footnote{http://archive.spitzer.caltech.edu/},
 we also collected the spectra of \emph{Spitzer}'s IR Spectrograph (IRS).  

\begin{figure*}
\includegraphics[width=150mm]{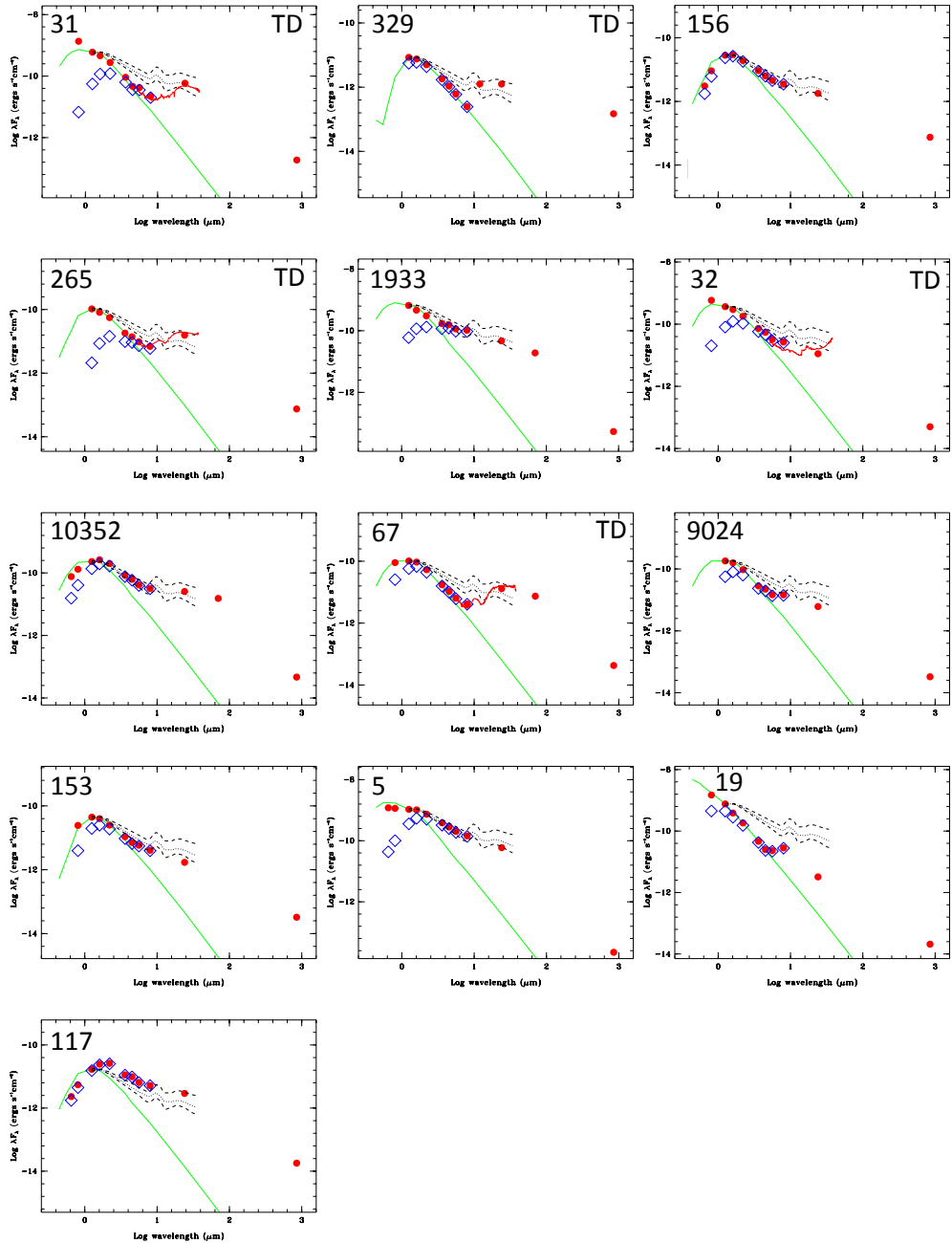}
\caption{The Spectral Energy Distributions of the 13 discs detected in our SCUBA-2 map ordered by decreasing  850 $\mu$m  flux.  Five objects with transition disc SEDs are labeled  ``TD" and are among the brightest eight sources. The green lines are the stellar photospheres. 
The blue boxes represent the observed optical and IR photometry before correcting for extinction.  The dotted lines correspond to the median mid-IR SED of K5-M2 CTTSs calculated by Furlan et al. (2006). The dashed lines are the quartiles.}
\label{fig:SEDs}
\end{figure*}

The resulting SEDs for the 13 discs detected by our SCUBA-2 observations are shown in Figure~\ref{fig:SEDs} ordered by decreasing 
850 $\mu$m flux.  
 One object, $\#$329 has no 24 $\mu$m detection in the  \emph{Spitzer} catalogs. Instead, we plot  their 12 and 22 $\mu$m fluxes from
 the Wide-field IR Survey Explorer (WISE)  catalog\footnote{http://irsa.ipac.caltech.edu/Missions/wise.html}  (9.52$\pm$0.16 mag and 7.35$\pm$0.16 mag, respectively). 
We find that five of the eight discs that are brightest at 850 $\mu$m  (sources $\#$ 31, 329, 265, 32, and 67, M$_D$  $>$ 3.0 M$_{JUP}$) have the mid-IR ``dip" that indicates the presence of an inner hole or a gap, which are the defining features of transitions discs. 
The SEDs of the 7 sources that were detected by the SMA at 1.3 mm (Lee et al. 2011) but remained undetected in our SCUBA-2 maps are not shown here, but they all have  ``normal" SEDs close to the median of Classical T Tauri stars (see their Figure 4). They also have lower disc masses than any of the transition discs in our sample (M$_D$ $<$ 3 M$_{JUP}$).

The correlation between a large 850 $\mu$m flux and a transition disc  SED seems to be robust. We note that  3 of these 5 transition discs (sources $\#$ 31, 32, and 67) have already been detected with the SMA (Lee et al.  2011, Espaillat et al. 2012) and that  object $\#$265 has a vey secure SCUBA-2 detection (S/N $\sim$ 14). Source $\#$329 is bright, but slightly elongated toward the southwest.  Still, we find no other 2MASS, WISE, or \emph{Spitzer} object  within 20$\arcsec$ of this source and consider the 850 $\mu$m detection to be associated to this object, although cloud contamination can not be ruled out.

\begin{figure*}
\includegraphics[width=150mm]{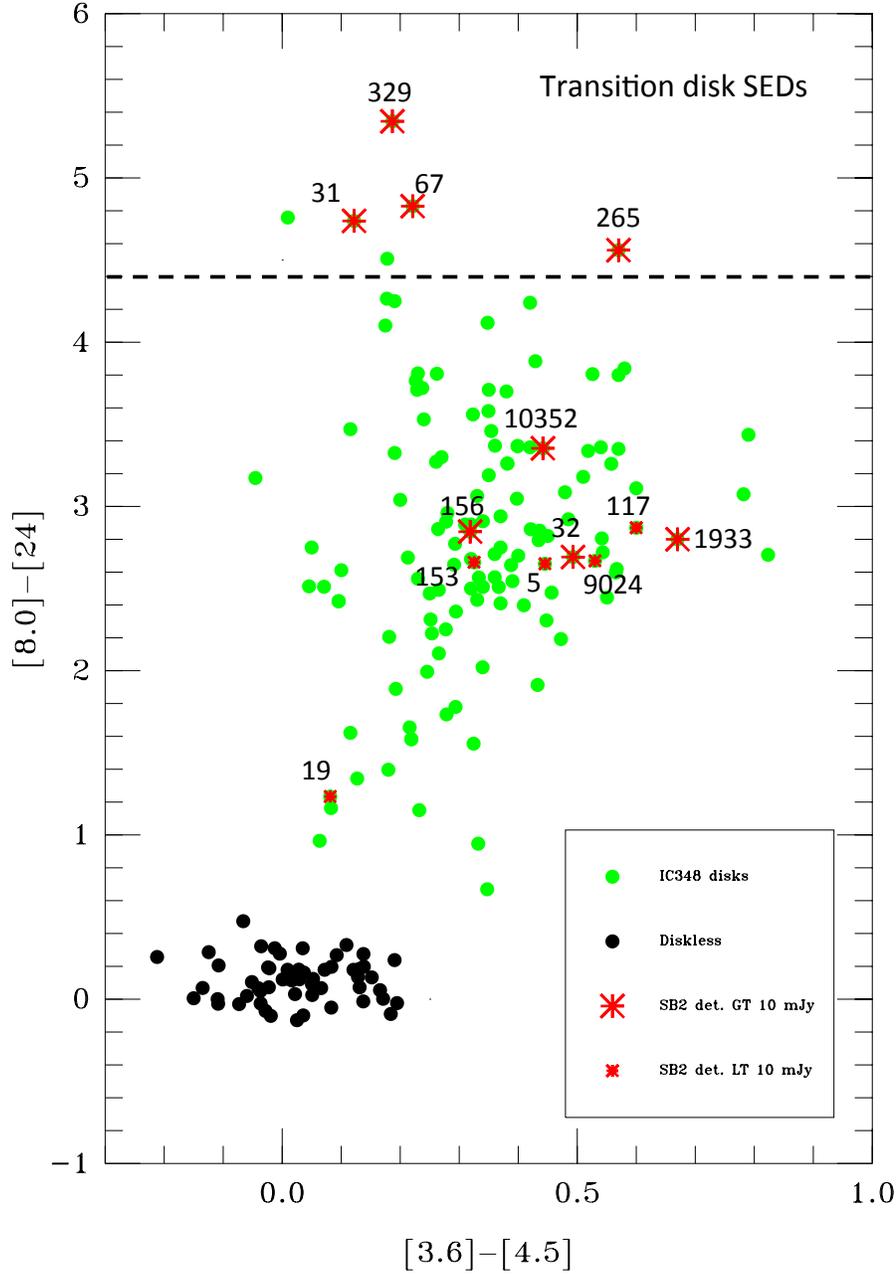}
\caption{The \emph{Spitzer}   [8.0]--[24] versus [3.6]--[4.5]  color-color diagram for discs in IC~348 (thick and anemic discs from L06 and Class ~II objects from M07). Our 850 $\mu$m detections are individually labeled. 
discs-less weak-line T Tauri stars from Cieza et al.  (2007) are shown  to illustrate the locus of bare stellar photospheres.  Transition objects have very red  [8.0]--[24]  colors as their SEDs that are steeply rising in the mid-IR, indicating the presence of  inner holes in their discs.
There is a clear correlation in our sample between a transition disc SED and a bright 850 $\mu$m flux. Four out of the six discs with the reddest
[8.0]--[24] colors are  among the brightest SCUBA-2 detections (850 $\mu$m flux $>$ 10 mJy).   
}
\label{fig:colors}
\end{figure*}

To establish how statistically significant this correlation is, we first need to estimate the number of transition discs present in IC~348. 
This is  a non-trivial task  as no standard definition exists for transition objects and their identification also depends on the available data. For instance, object $\#$ 32 can be identified as a transition disc based on the IRS spectra, but its broad-band photometry seems rather ``normal". 
To address this problem, we plot the [8.0--24] vs [3.6--4.5] color-color diagram (see Figure~\ref{fig:colors}) for our entire sample of discs (thick, anemic, and Class II sources).  Transition discs can be characterized by a steeply rising mid-IR SED (e.g., Espaillat et al.  2012).  The  [8.0--24] color nicely captures this feature, except for objects with very large inner holes.  In fact, all transition discs in our SCUBA-2 sample have 
[8.0--24] $>$ 4.4, except for object $\#$ 32.
From Figure~\ref{fig:colors}, we conclude that  66.7$\%$  (4/6) of the objects with [8.0--24] $>$ 4.4 have massive discs (M$_{D}$ $>$ 3 M$_{JUP}$), while only 3$\%$ (4/127)  of the  objects with  [8.0--24] $<$ 4.4 do so.  
To further quantify the significance of this correlation, we perform a Kolmogorov-Smirnov test to estimate the probability that the 
[8.0--24] colors of the discs with masses smaller and larger than 3 M$_{JUP}$ are drawn from the same parent population. We find that
the  Kolmogorov-Smirnov statistic,  $d$ (the maximum deviation between the cumulative   distributions),  is 0.48,  and that the probability of the two  distributions been drawn from the same parent population is 3$\%$.  The modest significance of the test for the given $d$ number is due to the small size of the transition disc sample.

A similar connection between massive discs and transition disc SEDs is found in other regions. For instance, in $\sigma$~Ori, Williams  et al. (2013) also  noted that 3 out of the 8 discs detected at 850 $\mu$m are transition objects.   Furthermore, the most massive discs in older regions such  as  Upper Scorpius and the TW Hydra Association  are also transition objects: [PZ99] J160421.7-213028 (Mathews et al. 2012)  and TW Hydra itself  (Calvet et al. 2002). 

The origin of transition discs is still a matter of intense debate (see Espaillat et al. 2014 for a recent review),  and several mechanisms have been proposed to explain their inner holes, including grain growth (Dullemond $\&$ Dominik, 2005), photoevaporation (Gorti $\&$ Hollenbach 2009), and the dynamical interaction with (sub)stellar objects (Quillen et al. 2004).
However, grain growth alone has been shown to be an inefficient hole formation mechanism as dust fragmentation and radial drift result in the efficient replenishment of micron-size grains (Brauer, Dullemond $\&$ Henning 2008; Birnstiel et al. 2012). Also, while a few transition discs such as CoKu Tau/4 were found do be close stellar binaries (Ireland $\&$ Kraus, 2008), it soon became clear that most transition discs are not due to binarity  (Pott et al. 2010). 
There seem to be  distinct populations of transition objects (Owen $\&$ Clarke,  2012), and massive discs with large inner holes are believed to be carved by recently formed massive planets (Najita et at. 2007; Cieza et al. 2012;  Kraus $\&$ Ireland, 2012;  Kim et al. 2013). 

In this context, we speculate that the correlation between a large disc mass and a transition disc SED  could be an initial conditions effect. 
Recent statistics on extra-solar planets imply that the usual outcome of disc evolution is a planetary system (Howard et al. 2010).  However, massive planets are relatively rare (Fressin et al. 2013) and not all discs will form planets massive enough to open wide gaps in the disc (Zhu et al. 2011; Dodson-Robinson $\&$ Salyk, 2011).  More massive discs  will naturally tend to form more massive planets (Mordasini et al. 2012), which in turn  will result in transition disc SEDs.
We also note that disc evolution coupled to planet formation \emph{could} also reenforce this trend. Najita et al. (2007) showed that transition discs tend to have stellar accretion rates $\sim$10 times lower and median disc masses  $\sim$4 times larger than non-transition objects. 
Massive enough planets could reduce or stop  accretion onto the star by isolating the inner and outer discs (Lubow et al. 1999); therefore,  the formation of one or more  massive planets  could result in both a transition disc SED and a reduction in the accretion rate, increasing the lifetime of the outer disc.  
However, the formation of a gap by a massive planet may also have the opposite effect on the lifetime of the outer disc as the inner edge of the gap could be directly exposed to stellar radiation and photoevaporate more efficiently (Owen et al. 2011). 
The net result of the formation of a gap by a massive planet (an increase versus a decrease in the lifetime of the outer disc) will depend on whether photoevaporation rates can approach the accretion rates of T Tauri stars, which is also still a matter of intense debate (see Alexander et al. 
2014 for a recent review). 

\subsection{Comparison to other regions:}

Investigating  the evolution of disc masses as a function of stellar age is of critical importance to planet formation theory as it provides a first-oder approximation of 
the amount of raw material still available for planet building. 
However, comparing disc mass  distributions between regions of  different ages is complicated by low  detection rates at millimeter wavelengths (Williams et al. 2013; Dods et al. 2015) and the strong dependence of the  millimeter wavelength  luminosity on stellar  mass (L$_{mm}$ $\propto$ M$^{1.5-2.0}$;  Andrews et al. 2013). 
To take into account the host mass effect, we follow the methodology introduced in Andrews et al. (2013) to compare the disc masses in IC~348 with those in Taurus. 
Taurus is a well-studied region for which we have the disc mass distribution for each spectral type.
Nevertheless,  even in Taurus the millimeter detection rate is a strong function of spectral type: it is close to 100$\%$ for stars K6 and earlier, but drops to 
$\lesssim$30$\%$ for  M3.5 to M6-type stars.  

Based on the spectral type of the host stars in IC~348, we can associate their stellar mass to the disc properties observed in Taurus. 
Thus, using the spectral type of all IC~348 discs, we can generate many sets of random ensembles based on Taurus disc luminosity distributions (see Figure 9, left panel). 
For this study, we simulate 10$^5$ synthetic Taurus disc ensembles that resembles to IC~348 in terms of  their stellar spectral types (i.e., dominated by M3 to M6-type stars). 
%
%
Then,  using the two-sample test for censored datasets (Feigelson $\&$ Nelson 1985), we compare the actual SCUBA-2
observations with the simulated Taurus ensembles to find the probability that they are not originated from the same parent distribution, p$_{\phi}$.
The cumulative probability distribution,  f ($<  p_{\phi}$), is presented in Figure 9 (right panel) and indicates that there is a  5$\%$ chance that the  850 $\mu$m disc luminosity  distribution in IC~348 is different from Taurus at $>$ 2-$\sigma$ significance. 
For comparison,  we plot the results of the same analysis for the  $\rho$ Ophiuchus star-forming region and Upper Scorpius from Andrews et al. (2013) and $\sigma$-Orionis from Williams et al. (2013). 

Given the available data, only the disc luminosity distribution of $\sigma$-Orionis (age $\sim$3-5 Myr, IR disc fraction $\sim$27$\%$; Williams et al. 2013) is significant different ($>$3-$\sigma$) from that of Taurus (age $\sim$1-3 Myr,  IR disc fraction $\sim$67$\%$; Rebull et al. 2010);  nevertheless, we emphasize that these statistical tests are based on very incomplete data sets and should be interpreted with caution.  
PMS star ages are very uncertain, but  IC~348 has an IR disc fraction of $\sim$50$\%$ and an estimated  age of 2-3 Myr (Lada et al. 2006),  and thus is particularly useful to establish the distribution of disc masses at the time half the disc population has  already dissipated.
However, we conclude that much deeper  millimeter wavelength surveys are needed to investigate disc mass distributions as a function of cluster age in detail.

On the other hand, current millimeter surveys do provide valuable information on the incidence of very massive discs in young stellar cluster. 
As discussed by Dodds et al. (2015),  discs exceeding the canonical Minimum Mass Solar Nebula (MMSN),  ranging from 10 to 100 Jupiter 
masses (Weidenschilling 1977),  are relative rare ($\lesssim$10-20$\%$) in young clusters.  
Our results in IC~348 strengthen  this conclusion since only  2 out  of 370 ($\sim$0.5$\%$) cluster members in our SCUBA-2 field (objects  $\#$ 31 and  329)
have disc masses larger than 10 M$_{JUP}$.  The fact that massive discs are rare,  \emph{specially} around very low-mass stars, is in agreement with both  the lower incidence of  giant planets compared to terrestrial planets (Ford 2014) and the lower fraction of M-type stars hosting Jovian planets with respect to solar-type stars (Gaudi et al. 2014).

\section[]{Conclusions }

We have observed  $\sim$370 members of the 2-3 Myr cluster IC~348. 
We have detected a total of 13 discs, raising the total number of detections in the cluster to 20. 
We have also detected 8 Class 0/I protostars. 
The detected discs have inferred masses ranging from 1.5 to 16 M$_{JUP}$.
Our results imply that disc masses exceeding the MMSN are very rare ($\lesssim$1$\%$) at the age of IC~348, specially around very low-mass stars. 
In our sample, we find a strong correlation between a relatively massive disc (M$_D$ $>$ 3 M$_{JUP}$) and 
a transition disc SED. 
We suggest  that this could be an initial conditions effect (e.g., more massive discs tend to form giant planets that result in transition disc SEDs) and/or a disc evolution effect (the formation of one or more  massive planets  results in both a transition disc SED and a reduction of the accretion rate, increasing the lifetime of the outer disc). 
Resolved ALMA images of cluster members could be used to study the structure of these transition discs in detail and  to identified additional objects with 
large  cavities  among the population of discs that  present normal IR SED (i.e., transition objects with partially depleted cavities). 
A stacking analysis of the Class II/Thick discs that individually remain undetected at 850 $\mu$m suggests that their median dust mass is 
$\lesssim$1 M$_{\oplus}$, corresponding to M$_{D}$ $\lesssim$ 0.3 M$_{JUP}$ if a gas to dust mass ratio of 100 is assumed. 
The IC~348 cluster represents an ideal target to investigate the distribution of disc masses at the 2-3 Myr age range, but the available data is not
deep enough to allow a meaningful comparison to older and younger regions. 
While the estimated  median flux  of the discs in the cluster ($\lesssim$1 mJy at 850 $\mu$m) is well beyond the sensitivity of 
our SCUBA-2 study, we note that such sensitivity level (5-$\sigma)$ could be reached in $\sim$1 min of integration time 
per object with the Atacama Large Millimeter/submillimeter Array (ALMA). 
Spectroscopic observations with ALMA would also be able to establish the gas content of the discs to investigate the evolution of the gas to dust mass ratio, while multi-wavelength photometry (e.g., at  850 $\mu$m and 3 mm) could be used to trace the growth of grains up to cm sizes. 
Such studies in clusters of different ages will eventually provide a deeper understanding  of the planet formation timescales than the \emph{Spitzer} plots of the IR disc fractions as a function of cluster age that are currently used to constrain the time available for planet formation.

\section*{Acknowledgments}
L.A.C. was supported by ALMA-CONICYT grant number 31120009 and CONICYT-FONDECYT grant number 1140109. L.A.C., M.S.,  and S.C. acknowledge support from the Millennium Science Initiative (Chilean Ministry of Economy),  through grant ÒNucleus P10-022-FÓ.

\begin{table*}
 \centering
 \begin{minipage}{140mm}
  \caption{Observing Log}
  \label{table:log}
  \begin{tabular}{@{}llrrrrlrlr@{}}
  \hline
  Date     &   Pogram-ID  & $\tau_{225}$ & T$_{int}$\\
                &                          &                         &  (hs)          \\
 \hline
2011-10-24            & M11BH29B       &   0.06 &  0.4 \\
2011-10-25            & M11BH29B        &  0.09 & 1.4\\
2011-11-10            & M11BH29B        &  0.13  & 1.4 \\
2011-12-17             & M11BH29B        & 0.11  & 1.5\\      
2012-02-03             &  M12AH29B       & 0.07  &  0.8    \\  
2012-08-24             &  M12BH10C       & 0.08  & 1.4    \\ 
2012-08-25              &   M12BH10C     & 0.06  &  2.1   \\
2012-08-26              &   M12BH10C     & 0.06  &  0.7   \\
2012-08-27              &  M12BH10C      & 0.08  &  1.4   \\
2013-01-07              &  M12BH10C      & 0.09  &   0.8   \\ 
2013-09-21              &  M13BH12A       & 0.05  & 0.5 \\
2013-09-23              &  M13BH12A       & 0.11  & 3.0 \\
2013-09-24              &  M13BH12A       &  0.09 & 3.5 \\
2013-10-29              &  M13BH12A       &  0.10 & 0.5\\
2013-10-30              &  M13BH12A       &  0.13  & 3.0\\
\hline
TOTAL   &                     &                              &   22.4     \\
\hline
\end{tabular}
\end{minipage}
\end{table*}

\begin{table*}
 \centering
 \begin{minipage}{140mm}
  \caption{Detected discs  (sorted by declining disc mass)}
  \label{table:detections}
  \begin{tabular}{@{}rccllccclr@{}}
  \hline
  ID   &  Ra (J2000)& Dec (J2000)& SpT  &  disc Class &          F$_{850}$               &   F$_{860}$          & F$_{1300}$ &    EW$_{H\alpha}$               &     M$_{disc}$\\
          &    (deg)       & (deg)             &            &            &           (mJy)                      &   (mJy)                    &   (mJy)            &      (\AA)                                 &   (M$_{JUP}$)   \\                                  
 (1)    &   (1)             &    (1)               &  (1)    &   (2)     &               (3)                       &    (4)                        &   (5)                 &   (1)                                        &       (6)          \\
 \hline
31      & 56.07583  &  32.08250   &   G1  &       ANEMIC/TD      &          52.9$\pm$4.0        & 62$\pm$6.0                &           $......$                     &          11$\pm$0.5            &          16.0          \\  
329    & 56.06490  & 32.15609    &   M7.5 &     STAR/TD        &           42.0$\pm$11        &      $......$                          &        $......$                               &       $......$                           &          12.6     \\      
156    & 56.02833  &  32.13172   &  M4   &       THICK/II             &           21.2$\pm$1.7        &      $......$                                 &        $......$                           &          10.0$\pm$2            &           6.4            \\ 
265    & 56.14458  &  32.26666   &  M3.5 &     II/TD                    &          21.1$\pm$1.6          &      $......$                          &        $......$                               &       $......$                           &          6.4     \\
1933  & 56.31812  & 32.10553    &  K5    &      II                          &         15.2$\pm$5.0          &      $......$                          &        $......$                               &       $......$                           &          4.6     \\    
   32    & 56.15792  &  32.13450   &  K7   &        THICK/TD        &          14.2$\pm$2.0         &       $......$                              &  5.50$\pm$0.80    &           68$\pm$1              &           4.3           \\
10352&  56.33525  &  32.10956  &  M1  &        THICK/II            &           13.2$\pm$2.9       &    $......$                              &    11.5$\pm$1.0    &            $......$                  &           4.0                  \\  
  67     & 55.93583   &  32.13831  &  M0   &       ANEMIC/TD     &          12.1$\pm$1.7        &  25$\pm$11        &         $......$                                  &          35$\pm$2                      &          3.6            \\
221     &   56.16772 &	 32.15923    &   M4.5   &   THICK/II       &                $......$               &         $......$              &  3.23$\pm$0.89   &         40$\pm$5                             &            2.9   \\
 9024  & 56.14739  &  32.12671   &   M0    &   THICK/II               &         9.3$\pm$1.5          &          $......$                               &           $......$                            &       $......$                           &                        2.8         \\   
153  & 56.17822  &  32.14273      & M4.75  &   THICK/II         &          9.2$\pm$2.8          &         $......$                              &   6.49$\pm$0.71 &   40$\pm$10            &                      2.8           \\
 248  &    56.14978  & 32.15674    &   M5.25 &   THICK/II        &              $......$                 &        $......$                  &   2.81$\pm$0.81     &        30$\pm$5                            &            2.6   \\  
468  &    56.04613  & 32.02880    &   M8.25  &    THICK/II     &               $......$                    &     $......$                 &   2.37$\pm$0.56      &      400$\pm$50                          &            2.2   \\  
      5 &    56.10844  & 32.07511     &   G8        &   THICK/II     &               6.3$\pm$1.8       &       $......$                &    $......$                     &       $......$                                     &            1.9          \\  
8078&    56.11120&  32.13897     &   M0.5     &    THICK/II     &               $......$                &      $......$                   &   2.07$\pm$0.51      &        75                                          &            1.9    \\
100  &     56.09301& 32.20021     &   M1        &    THICK/II    &                  $......$              &       $......$                   &   2.06$\pm$0.51      &        90                                          &           1.9     \\ 
192  &      56.09850& 32.03130    &    M4.5   &    THICK/II       &               $......$             &       $......$                     &   2.06$\pm$0.57     &       40$\pm$20                         &            1.9     \\
19        &  56.12842  &  32.16550      &  A2       &     THICK/II        &           5.9$\pm$1.7        &       $......$             &    $......$                     &       $......$                                &                        1.8          \\ 
15   &       56.18631 & 32.06742   &    M0.5   &    THICK/II          &                $......$             &    $......$                                   &  1.98$\pm$0.46      &      36                                          &            1.8     \\
117    &    55.99617&   32.23925     &    M3      &       II                 &              5.1$\pm$1.3       &       $......$                                   &    $......$         &       $......$                                  &                       1.5          \\   
\hline
\end{tabular}
NOTES: (1) Source ID, coordinates, spectral types and H$\alpha$ EWs from Luhman et al.  (2003) or Muench et al.  (2007). (2) Object type from Lada et al. (2006) or Muech et al.  (2007)/ this paper. (3) 850 $\mu$m flux from our SCUBA-2 data.
(4) 860 $\mu$m flux from SMA data (Espaillat et al. 2012). (5) 1300 $\mu$m flux from SMA  (Lee, Williams $\&$  Cieza, 2011).  (6) disc mass derived as described in Section~\ref{disc-mass} for SCUBA-2 detections, or taken from  Lee, Williams $\&$  Cieza (2011) for objects detected only at 1300 $\mu$m.
\end{minipage}
\end{table*}

\begin{table*}
 \centering
 \begin{minipage}{140mm}
  \caption{Detected Protostars (sorted by declining flux)}
 \label{table:proto}
  \begin{tabular}{@{}rccclccclr@{}}
  \hline
  ID   &  Ra (J2000)& Dec (J2000)&  Object  Class &          F$_{850}$                  \\
          &    (deg)       & (deg)             &                            &           (mJy)                          \\
 (1)    &   (1)             &    (1)               &  (1)                    &                                               \\
  \hline
57025 &  55.98704  &  32.05094   &       0                     &         756$\pm$85          \\
HH221&  55.98654 &  32.01386  &        0                     &         727$\pm$98          \\   
 1898  &  56.18288	& 32.02704   &        0/I                  &         451$\pm$23           \\
54362 &  55.96228  & 32.05729    &        I                      &        189$\pm$33           \\
 51       &  56.05417  &  32.02650   &        I                     &         129$\pm$10           \\
 904     &  56.30754  & 32.20281    &        I                     &           19$\pm$1.4          \\
 \hline
\end{tabular}

NOTES: (1) Source ID, coordinates, and object type from Luhman et al.  (2003) or Muench et al.  (2007).
\end{minipage}
\end{table*}

\begin{table*}
 \centering
\begin{minipage}{135mm}
  \caption{Stacking photometry}
  \label{table:stacking}
  \begin{tabular}{@{}lrrr@{}}
  \hline
Sample     &    Sample   &  average &  median      \\ 
                   &     Size        &    (mJy)  &   (mJy)      \\             
\hline
L06     discless                      &  134   &    0.46$\pm$0.23  &   0.50$\pm$0.25               \\
L06     Anemic                       &     67  &    -0.12$\pm$0.28     & -0.19$\pm$0.30             \\
\textbf{L06     Thick}                            &    \textbf{83}    &       \textbf{0.79$\pm$0.28}   &  \textbf{1.01$\pm$0.30}                 \\
L06     Thick +  Anemic         &    150   &   0.30$\pm$0.22        &     0.35$\pm$0.25           \\
M07    Class II                        &     35 &    0.78$\pm$0.59   &        0.95$\pm$0.61               \\
\textbf{M07    Class II +  L06 Thick} &   \textbf{118} &  \textbf{0.69$\pm$0.26}        &   \textbf{1.02$\pm$0.29}                  \\     
\hline
\end{tabular}
\end{minipage}
\end{table*}

\begin{table*}
 \centering
 \begin{minipage}{140mm}
  \caption{Optical and IR photometry for detected discs  (sorted by declining disc mass)}
  \label{table:detections}
  \begin{tabular}{@{}rrrrrrrrrrrr@{}}
  \hline
 ID   &   R-band   &   I-band & J-band                 &   H-band            &  K-band                      &  F$_{3.6}$ &          F$_{4.5}$               &   F$_{5.8}$          & F$_{8.0}$ &    F$_{24}$  & F$_{70}$ \\
        &    (mag)     &   (mag)  &   (mJy)                  & (mJy)                   &  (mJy)                        &    (mJy)      &           (mJy)                     &       (mJy)               &   (mJy)      &   (mJy)          &  (mJy)  \\
 \hline
   31    & ......  & 15.37  &    23.20  &  62.20  &   88.80  &   75.59  &   54.10  &   66.20  &  53.40  &  469.00  &    .......  \\
  329    &  20.17  & 17.64  &     2.32  &   3.16  &    3.19  &    2.04  &    1.55  &    1.11  &   0.65  &    ....... &    .......  \\
  156    &  17.27  & 15.48  &     9.83  &  14.20  &   13.30  &   10.90  &    9.35  &    8.46  &   9.30  &   14.30  &    ....... \\
  265    & ......  & ...... &     0.88  &   4.58  &   10.60  &   11.80  &   13.80  &   13.60  &  15.90  &  125.00  &    ....... \\
 1933    & ......  & ...... &    25.00  &  62.00  &   97.30  &  140.00  &  183.00  &  180.00  & 252.00  &  373.00  &    445 \\
   32    & ......  & 14.18  &    33.70  &  65.90  &   77.90  &   69.10  &   69.60  &   54.10  &  67.00  &   89.40  &    ....... \\
10352    &  14.90  & 13.42  &    57.00  & 104.00  &  122.00  &   92.10  &   88.50  &   74.20  &  83.10  &  204.00  &     357 \\
   67    & ......  & 13.93  &    24.10  &  36.09  &   32.09  &   19.00  &   14.90  &   11.40  &  10.90  &  104.00  &     173  \\
 9024    & ......  & ...... &    23.50  &  43.00  &   46.40  &   28.30  &   29.50  &   25.30  &  37.30  &   48.70  &    ....... \\
 153     & ......  & 15.95  &     8.18  &  13.30  &   13.80  &   11.40  &    9.84  &   10.30  &  10.50  &   13.60  &    ....... \\
   5     &  13.79  & 12.44  &   149.00  & 291.00  &  371.00  &  389.00  &  375.00  &  349.00  & 369.00  &  474.00 &    .......  \\
  19     & .....   & 10.81  &  186.00   & 151.00  &  112.00  &   51.20  &   35.29  &   41.20  &  74.09  &   25.80 &    .......    \\
 117     & 17.27   & 15.82  &    6.42   &  12.50  &   18.60  &   13.30  &   14.60  &   12.00  &  13.70  &   23.30   &    ....... \\
\hline
\end{tabular}
NOTES: Optical Photometry comes from Luhman et al. 2003, Littlefair et al. 2005,  Cieza $\&$ Baliber, 2006, and  Cieza et al. 2007)
 comes from \emph{Spitzer's Cores to Discs} catalog (Evans et al. 2009).
\end{minipage}
\end{table*}

\begin{figure*}
\includegraphics[width=80mm]{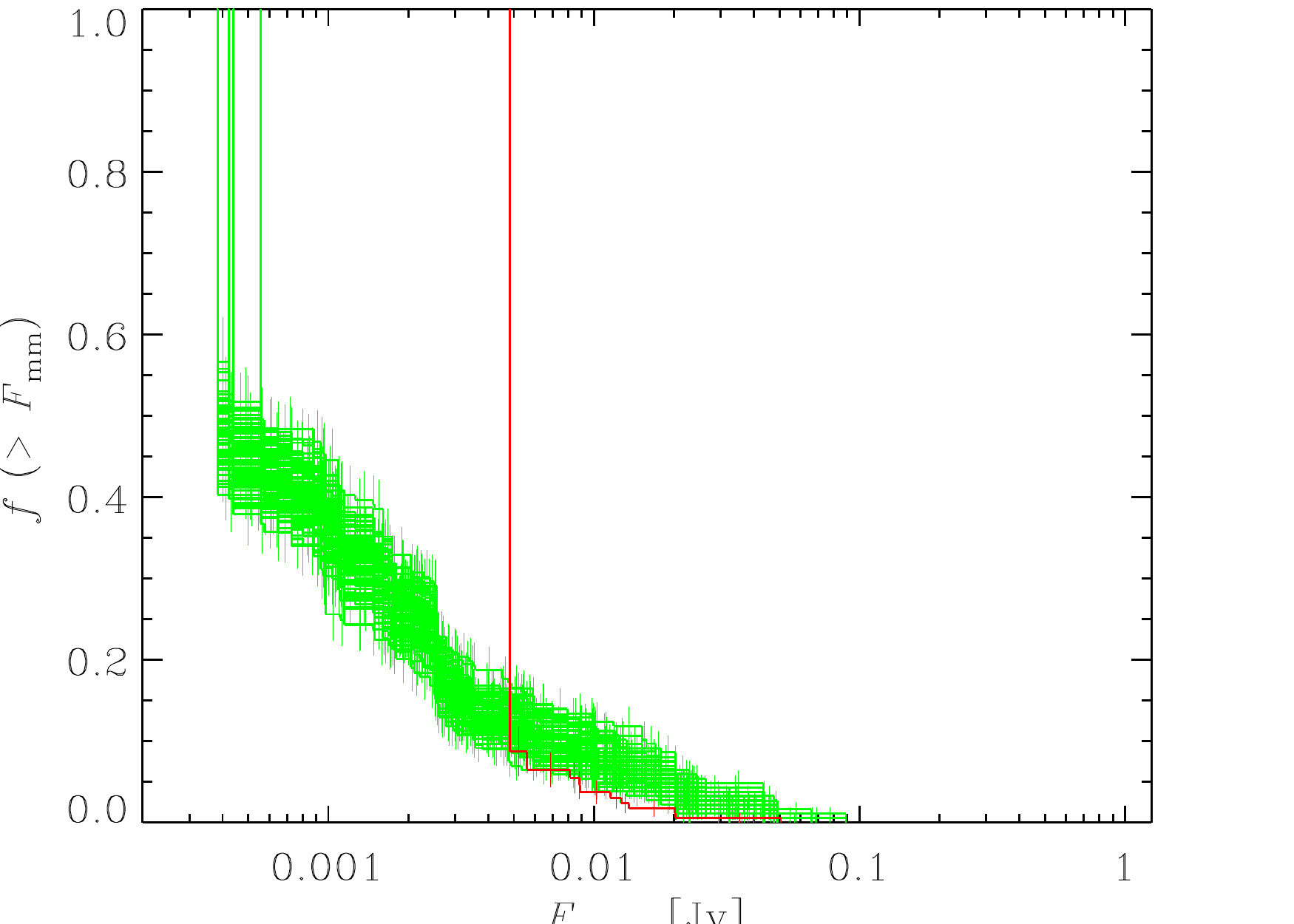}
\includegraphics[width=70mm]{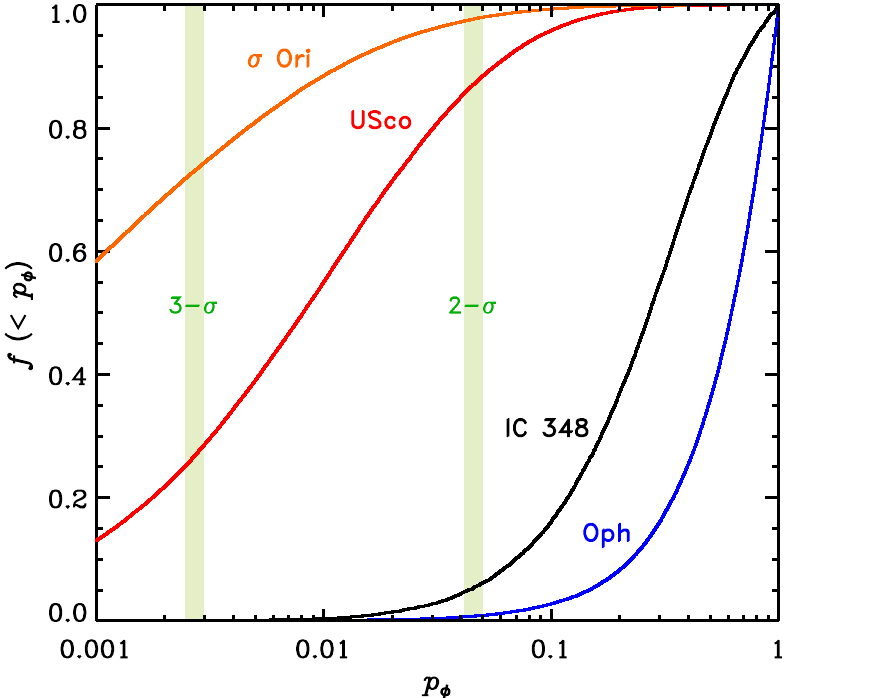}
\caption{\textbf{Left:} The cumulative distribution function (CDF) of  the disc luminosities in IC~348 (red) compared to 100 random draws of  the same CDF  for Taurus with the same spectral type distribution as in IC~348 (green). The Taurus distribution has been scaled down to compensate for its closer distance (140 pc vs 320 pc).  The fluxes in IC~348 tend to be slightly lower than in Taurus, but this difference is not significant given the available data.
\textbf{Right:} The comparison between the disc luminosity distribution of IC~348 and Taurus.
The quantity $p_\phi$ is  the probability that two samples statistically belong to the same population. 
The cumulative distribution of   $p_\phi$  is calculated for 10$^5$ Monte Carlo simulations. 
Our results for the IC~348 disc luminosity show no significant discrepancy with those of Taurus. 
As a reference, similar comparisons have been made for the millimeter surveys of discs in $\rho$ Ophiuchus and
Upper Scorpius from Andrews et al. (2013), and $\sigma$ Orionis from Williams et al. (2013).} 
\label{fig:comparison}
\end{figure*}

\bsp

\label{lastpage}

\end{document}